\documentclass[twocolumn]{aa}
\usepackage{graphicx}
\usepackage{times}
\usepackage{natbib}
\bibpunct{(}{)}{;}{a}{}{,}

\usepackage{color}

\usepackage{amssymb}
\begin{document}

  \title{Probing the atmosphere of the bulge G5III star OGLE-2002-BUL-069 by analysis of microlensed $\element{H}\alpha$ line
    \thanks{Based on observations made at ESO, 69.D-0261(A), 269.D-5042(A), 169.C-0510(A)  }}
  \titlerunning{Probing the atmosphere of a bulge G5III star}
  \authorrunning{ A.~Cassan, J.P.~Beaulieu, S.~Brillant et al.}
  \author{ 
    A.~Cassan\inst{1,2},              J.P.~Beaulieu\inst{1,2},
    S.~Brillant\inst{1,3},              C.~Coutures\inst{1,2,4},
    M.~Dominik\inst{1,5},
    J.~Donatowicz\inst{1,6},            U.G.~J{\o}rgensen\inst{1,7},
    D.~Kubas\inst{1,8},               M.D.~Albrow\inst{1,9}, 
    J.A.R.~Caldwell\inst{1,10},
    P.~Fouqu\'e\inst{1,11},           J.~Greenhill\inst{1,12},
    K.~Hill\inst{1,12},               K.~Horne\inst{1,5},
    S.~Kane\inst{1,5},                R.~Martin\inst{1,13},
    J.~Menzies\inst{1,14},            K.R.~Pollard\inst{1,9},
    K.C.~Sahu\inst{1,10},             C.~Vinter\inst{7}, 
    J.~Wambsganss\inst{1,8},          R.~Watson\inst{1,12},
    A.~Williams\inst{1,13}
    \and C.~Fendt\inst{8}        \and P.~Hauschildt\inst{15}
    \and J.~Heinmueller\inst{8}  \and J. B.~Marquette\inst{2}
    \and C.~Thurl\inst{16}
  }
  \offprints{beaulieu@iap.fr}
  \institute{
{PLANET collaboration member} 
\and
{Institut d'Astrophysique de Paris, 98bis Boulevard Arago, 75014 Paris, France} 
\and
{European Southern Observatory, Casilla 19001, Vitacura 19, Santiago, Chile} 
\and
{DSM/DAPNIA, CEA Saclay, 91191 Gif-sur-Yvette cedex, France} 
\and
{University of St Andrews, School of Physics \& Astronomy, North Haugh, St Andrews, KY16~9SS, United Kingdom} 
\and
{Technical University of Vienna, Dept. of Computing, Wiedner Hauptstrasse 10, Vienna, Austria} 
\and
{Niels Bohr Institute, Astronomical Observatory, Juliane Maries Vej 30, DK-2100 Copenhagen, Denmark} 
\and
{Universit\"at Potsdam, Astrophysik, Am Neuen Palais 10, D-14469 Potsdam, Germany} 
\and
{University of Canterbury, Department of Physics \& Astronomy, Private Bag 4800, Christchurch, New Zealand } 
\and
{Space Telescope Science Institute, 3700 San Martin Drive, Baltimore, MD 21218, USA} 
\and
{Observatoire Midi-Pyrenees, UMR 5572, 14, avenue Edouard Belin, F-31400 Toulouse, France} 
\and
{University of Tasmania, Physics Department, GPO 252C, Hobart, Tasmania 7001, Australia} 
\and
{Perth Observatory, Walnut Road, Bickley, Perth 6076, Australia} 
\and
{South African Astronomical Observatory, P.O. Box 9 Observatory 7935, South Africa} 
\and
{Hamburger Sternwarte, Gojenbergsweg 112, 21029 Hamburg, Germany} 
\and
{RSAA, Mount Stromlo and Siding Spring Observatories, ANU, Cotter Road, Weston Creek, Canberra, ACT 2611, Australia} 
}
     
  \date{Received ; accepted}
   
  \abstract{ We discuss high-resolution, time-resolved spectra of the
caustic exit of the binary microlensing event OGLE 2002-BLG-069 obtained with UVES on the VLT. The
source star is a G5III giant in the Galactic Bulge. During such events, the source star is highly magnified, and a strong differential 
magnification around the caustic resolves its surface. Using an appropriate model
stellar atmosphere generated by the PHOENIX v2.6 code we obtain a model light
curve for the caustic exit and compare it with a dense set of photometric
observations obtained by the PLANET microlensing follow up network. 
We further compare predicted variations in the $\element{H}\alpha$ equivalent width with those
measured from our spectra. While the model and observations agree in the gross
features, there are discrepancies suggesting shortcomings in the model,
particularly for the $\element{H}\alpha$ line core, where we have detected
amplified emission from the stellar chromosphere after the source star's
trailing limb exited the caustic. This achievement became possible
by the provision of the very efficient OGLE-III Early Warning System, a network of small telescopes capable of
nearly-continuous round-the-clock photometric monitoring, on-line data reduction,
daily near-real-time modelling in order to predict caustic crossing parameters,
and a fast and efficient response of a 8m class telescope 
to a ``Target-Of-Opportunity'' observation request.
\keywords{Techniques: gravitational microlensing -- Techniques: high resolution spectra -- 
Techniques: high angular resolution -- Stars: atmosphere models -- Stars: individual (OGLE~2002-BLG-069)} } 
  \maketitle
%
\section{Introduction} \label{sec:intro}
%
 Near extended caustics produced by binary (or multiple) lenses,  the source
star undergoes a large total magnification in brightness. Furthermore, its
surface is differentially magnified because of a strong gradient in magnification. The 
relative lens-source proper motion is typically slow enough to allow the
light curve to be frequently sampled. This translates to a high spatial
resolution on the source star's surface and hence permits its radial brightness
profile to be inferred from the observations. Over the past four years,
coefficients characterizing linear or square-root limb-darkening profiles have
been obtained with the microlensing technique for several Bulge giants and
sub-giants (\citealt{Fields2003} and references therein) and a main sequence star \citep{Abe2003}. 
A new generation of stellar atmosphere models \citep{OH2000} have revealed limb-darkening 
laws that are significantly different from the traditional analytic ones \citep{Hendry2002a,claret03}.
The centre to limb variation of spectral lines can show markedly different
behaviour from that of the continuum. Most moderately strong and weak lines
weaken towards the limb, but resonant scattering lines can vary in a much more
pronounced way \citep{Dim95}. In cool giants, $\element{H}\beta$, being formed lower in the atmosphere,
is more limb-darkened than $\element{H}\alpha$. $\element{TiO}$ and  $\ion{Ca}{ii}$ show strong variations,
and some lines and bands may even be limb-brightened. Intensive spectroscopic
monitoring of a caustic exit at high resolution with high S/N should reveal
temporal variations in the equivalent widths of promising spectral lines that
can be compared with predictions from stellar atmosphere models \citep{HSL2000}.

 We present here the first photometric and spectroscopic monitoring campaign 
that has successfully been performed at high 
resolution with dense sampling.
Previously, \citet{Castro2001} obtained two KECK HIRES spectra of
EROS~2000-BLG-5, but missed the trailing limb  of the caustic where the effects are stronger, while
the data in the qualitative analysis of \citet{Albrow2001} involved
dense coverage but at low resolution. \citet{Afonso2001} provided a model that
reproduces the photometric data and found an excess of  $\element{H}\alpha$ at the limb.
They attribute the excess to chromospheric emission, but it could be due to shortcomings of the 
synthetic spectra.
   Using a determination of the spectral type of the source star, 
we have computed the limb-darkening profiles in $\mathrm{R}$ and $\mathrm{I}$ 
from appropriate PHOENIX synthetic spectra of the source star 
and fitted a fold-caustic model to our photometric data obtained during the caustic exit. 
This model has been used here to compute the synthetic spectrum for wavelengths around
the $\element{H}\alpha$-line.
A more detailed report on the determination of the stellar parameters and
the analysis of other spectral lines will be presented elsewhere
\citep{Beaulieu2004}, as will the details of a full binary lens model fit
to the PLANET observations of the event \citep{Kubas2004}.

%
\section{OGLE~2002-BLG-069 photometry and spectroscopy} \label{sec:photo}
%

The PLANET collaboration, comprising six different telescopes, namely, SAAO
1.0m (South Africa), Danish 1.54m (Chile), ESO 2.2m (Chile), Canopus 1.0m
(Australia), Mt. Stromlo 50in (Australia) and Perth 0.6m (Australia),
commenced photometric observations of OGLE 2002-BLG-069 alerted by the OGLE-III early Warning System \citep{Udal03}
in early June, 2002.
From online data reductions on 25 June it was realised that the event involved
a binary lens and a bright giant source star. With the source taking $\sim1.4$ days
to cross the caustic on entry, it appeared to be an excellent candidate for
time-resolved spectroscopy of the caustic exit. Using the predictions based on
modeling our photometry, and thanks to excellent coordination with the staff at
La Silla and Paranal, very good coverage was obtained during the caustic exit together with
post-caustic reference spectra. These observations were performed using the
UVES spectrograph mounted on Kueyen (VLT UT2) as part of Target-of-Opportunity and
Director's Discretionary time. Thirty-nine spectra with exposure time of 20 min (1 hour for the post caustic observations) 
were obtained alternately in the
so-called standard settings, 580 and 860, covering the full $4800-10600$~\AA{}
range at a resolution of 30000 with S/N ranging from 50 to 130.
   From an analysis of the curve-of-growth of 100 \ion{Fe}{i} and \ion{Fe}{ii} lines
as done by \citet{Minniti02}, and
independently from the study of the \ion{Ca}{ii} and \element{Mg} lines \citep{Jorgensen1992a}, a
good fit to all the data was obtained with a plane-parallel model atmosphere
having $T_\mathrm{eff} = 5000~$K, $\log(g) = 2.5$, $v_\mathrm{turb} = 1.5~\mbox{km}\,\mbox{s}^{-1}$ and
$[\element{Fe}/\element{H}] = -0.6$,
The calibration of the MK spectral type gives, for a G5III star,
 $T_\mathrm{eff}=5050~$K, $\mathrm{M_V}=+0.9$, $\mathrm{(V-I)}=0.95$,
 $\mathrm{M/M_\odot}=1.1$ and
  $R/R_\odot=10$.  At HJD=2452455.2754, the source was amplified by 17.7.
 The measured color from SAAO data are $I=13.01 \pm 0.01$, $(V-I)=2.06 \pm 0.02$, so
 $\mathrm{E(V-I)} = 1.11 \pm 0.02$. We adopt a conservative
 $\mathrm{A_V/E(V-I)=2.2\pm 0.3}$, and then derive a distance of $9.4 \pm 1.4~ \mathrm{kpc}$.

%
%


%
\section{Modeling of the caustic exit} \label{sec:model}
%
With $\rho$ denoting the fractional radius of the source star, we approximate
its wavelength-dependent brightness profile $I^\lambda(\rho) = \overline{I} \xi^{\lambda}(\rho)$
by concentric rings of constant intensity. A model atmosphere 
for the stellar parameters given in the previous section was computed 
with the PHOENIX grid v2.6 (based on the code described by \citet{Allard01} but using spherical
geometry and spherically symmetric radiative transfer). 
A synthetic spectrum was calculated from this model at a spectral
resolution of 0.05 \AA\ at 128 steps in source radius. The intensity profile was
interpolated with cubic splines at 1000 points equally spaced in
$\cos \vartheta = \sqrt{(1 - \rho^2)}$ where  $\vartheta$ is the
emergent angle, so that the width of the rings of constant
intensity decreases toward the stellar limb. For a broadband filter $(s)$, the 
intensity profile  $\xi^{(s)}(\rho)$ is obtained  by convolving the source brightness profile 
with the filter, CCD transmission  and atmosphere response functions.
%
%
The observed magnitude $m^{(s)}(t)$ reads  
$m^{(s)}(t) = m_\mathrm{S}^{(s)}-2.5 \lg \left\{A^{(s)}(t)+g^{(s)}\right\} $
where $m_\mathrm{S}^{(s)}$ is the  
intrinsic source magnitude,
$g^{(s)}=F_B^{(s)}/F_S^{(s)}$ is the ratio between background flux
$F_B^{(s)}$ and intrinsic source flux $F_S^{(s)}$, and $A^{(s)}(t)$ the source magnification
at time $t$.

\begin{figure}
\includegraphics[width=9.cm]{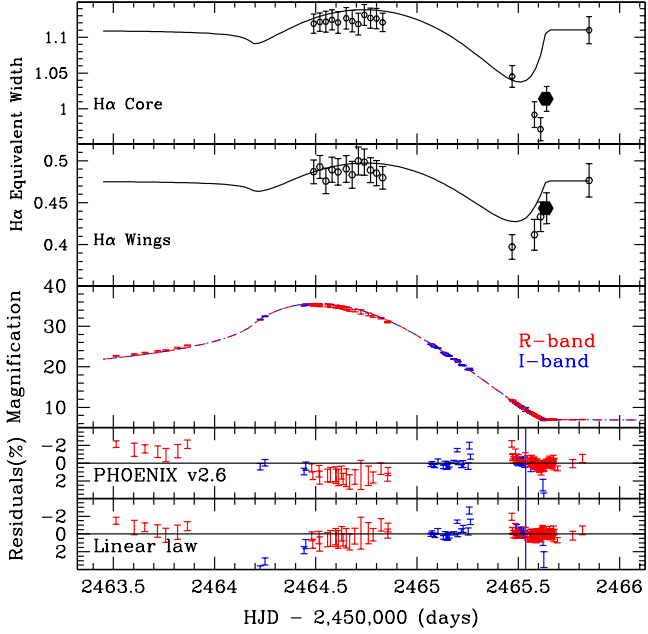}
\caption{
    Equivalent-width variation of the $\element{H}\alpha$ core (top panel) and wings (second panel from the top);
  the open circles in the plots of the equivalent-width variation
  correspond to our UVES data, while the adjoining solid lines represent the model
  predictions over the course of the caustic passage. The big dots correspond to the spectra of 10 July, UT 02H58.
    Third panel: model light curves and photometric data, where R-band data are plotted in red and I-band data in blue; 
  model residuals from the chosen PHOENIX atmosphere (fourth panel)
  and from the linear limb-darkening law (bottom panel), with the same color convention.
  The model parameters can be found in Tab. \ref{tab:fitparameters}. 
  The majority of model residuals are below the 2~\% level. The post caustic observations have been plotted at 2465.85,
but have been taken on 16 August.}
\label{fig:AetAcurve}
\end{figure}

The photometric data presented in the lower panel of Fig.~\ref{fig:AetAcurve} show a peak
with the characteristic shape resulting from a fold-caustic exit.
For a source in the vicinity of a fold caustic,
the magnification $A^{(s)}(t)$ 
can be decomposed \citep[e.g.,\ ][]{Albrow1999} as
$A^{(s)}(t) = A^{(s)}_\mathrm{crit}(t) + A_\mathrm{other}(t)$, where $A^{(s)}_\mathrm{crit}(t)$ denotes the 
magnification of the critical images associated with the caustic, and $A_\mathrm{other}(t)$ denotes
the magnification of the remaining images of the source under the action of the binary lens.

Let us consider a uniformly moving source crossing the caustic within 
$2\,\Delta t$ and exiting the caustic by its trailing
limb at $t_\mathrm{f}$.
If one neglects the curvature of the caustic and the variation of its strength over the 
size of the source, 
the magnification of the critical images 
reads
$A^{(s)}_\mathrm{crit}(t) = a_\mathrm{crit} \, G_\mathrm{f}\left(-\frac{t-t_\mathrm{f}}{\Delta t}\,;\,\xi^{(s)}\right)\,$
where $G_\mathrm{f}(\eta; \xi^{(s)})$ is a characteristic  function \citep{SW1987}
which solely depends on the intensity profile $\xi^{(s)}$. 
%
%
  For $|{\omega} (t- t_\mathrm{f})| \ll 1$,
the magnification of the non-critical images can be approximated by
$  A_\mathrm{other} (t) \simeq a_\mathrm{other}\left[1+  { \omega} (t- t_\mathrm{f})\right]$
  where $a_\mathrm{other}$ is the magnification at the caustic exit at time
$t_\mathrm{f}$ and ${\omega}$ measures its variation.

\begin{table}
  \caption[]{Model parameters of a  
fit to data obtained by PLANET with the SAAO~1.0m, UTas 1.0m, and ESO~2.2m.
on the microlensing event OGLE~2002-BLG-069.
While source magnitudes $m_\mathrm{S}^{(s)}$ and blend ratios $g^{(s)}$ result from a fit 
involving a binary lens model to the
complete data sets, 
the remaining parameters have been determined by applying
a generic fold-caustic model with PHOENIX limb darkening solely to the data taken around the caustic exit.}  
%
  \begin{center}
    \begin{tabular}{lr||lr}
      \hline Parameter & Value & Parameter & Value   \\ 
      \hline\rule{0pt}{2.8ex}
      $m_\mathrm{S}^{\mbox{\scriptsize SAAO~[$I$]}}$ & $16.05$ & $t_\mathrm{f}~\mathrm{(days)}$  & $2465.637$  \\
      \rule{0pt}{2.8ex}
      $m_\mathrm{S}^{\mbox{\scriptsize UTas~[$I$]}}$ & $16.05$ & $\Delta t~\mathrm{(days)}$ & $0.7297$  \\ 
      \rule{0pt}{2.8ex}
      $m_\mathrm{S}^{\mbox{\scriptsize ESO~2.2m~[$R$]}}$   & $16.3$  & $a_\mathrm{crit}$ & $19.60$ \\  
      \rule{0pt}{2.8ex}
      $g^{\mbox{\scriptsize SAAO~[$I$]}}$ & $0.16$ & $a_\mathrm{other}$  & $7.011$  \\
      \rule{0pt}{2.8ex}
      $g^{\mbox{\scriptsize UTas~[$I$]}}$ & $0.064$ &  
		${\omega}~\mathrm{(days)^{-1}}$ & $-0.04519$ \\ 
      \rule{0pt}{2.8ex}
      $g^{\mbox{\scriptsize ESO~2.2m~[$R$]}}$   & $0.0083$ \\
      \hline
    \end{tabular}
    \label{tab:fitparameters}
  \end{center}
\end{table}
As the period during which our generic fold-caustic model is believed to be a fair approximation, we
choose the range $2463.45 \leq \mbox{HJD}-2450000 \leq 2467$. Restricting our attention
to those data sets with more than two points in this region 
leaves us with 29 points from SAAO and 15 points from UTas in I as
well as 98 points from the ESO~2.2m in R, amounting to a total of 142~data points.
From a binary lens model of the complete data sets \citep{Kubas2004} for these observatories and filters,
we have determined the source magnitudes $m_\mathrm{S}^{(s)}$ and the 
blend ratios $g^{(s)}$.
With the adopted synthetic spectra, we compute the stellar intensity profiles
for I and R, and use these for obtaining a fit of the generic fold-caustic model 
to the data in the caustic-crossing region by
means of $\chi^2$-minimization, which determined the 5~model parameters 
$t_\mathrm{f}$, $\Delta t$, $a_\mathrm{crit}$, $a_\mathrm{other}$, and 
${\omega}$ as shown in Tab.~\ref{tab:fitparameters}. If a classical linear limb darkening is 
included in the list of parameters to fit (as in  \citealt{Albrow1999b}), 
the best fit is obtained with $\Gamma=0.5$.

 The modeled light curve and the structure  of the residuals of the two fits are
given in the lower panels of Fig.~\ref{fig:AetAcurve}. The fit with adopted PHOENIX limb darkenings 
clearly show systematic trends at the level of 1-2~\%.
The same trends in the residuals can be seen from the figure, which suggests some 
specific features lying in the real stellar atmosphere. 
  We checked whether this discrepancy can be caused by the straight fold caustic
 approximation, but from running the already mentioned global binary lens model
neither the very small curvature of the caustic nor the source trajectory can
explain this systematic effect.
\section{Comparison between synthetic spectra and spectroscopic data} \label{sec:compa}
\begin{figure}
\includegraphics[width=8.cm]{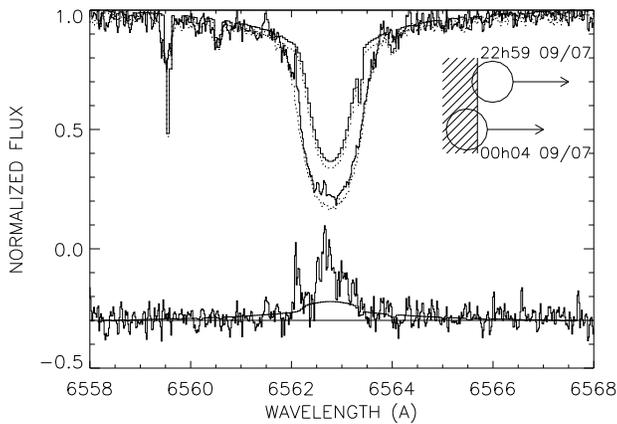}
\caption{
    Upper panel: two UVES spectra (two lower curves) corresponding to 9 July, UT 22h59 
  ($\mbox{HJD}-2450000 =2465.47$, solid line) and 9 July at UT 00h04 (reference spectra, 
$\mbox{HJD}-2450000 =2464.51$, dotted line),
  as well as two computed synthetic spectra at the same epochs (the two upper solid and dotted curves).
    Lower panel: fractional difference
  $\delta F^\lambda = 2 (F^\lambda_{00h04} - F^\lambda_{22h59})/(F^\lambda_{00h04} + F^\lambda_{22h59})$
  (lower solid line) for wavelengths in the vicinity of the $\element{H}\alpha$-line shifted vertically by -0.3. 
  Both the observations and the model are shown.
  On the upper right, we show the relative position of the source star at each epoch 
   with respect to the fold-caustic shown as dashed.}
\label{fig:ha}
\end{figure}

%
 The fold-caustic model derived in Sec. \ref{sec:model} is used to compute the source flux during the
caustic exit in order to obtain a synthetic spectrum at each epoch in a consistent way.
In Fig.~\ref{fig:ha} we show the observed and synthetic spectra at two epochs: 9 July at UT 00h04 
and 9 July at UT 22h59 (trailing limb crossing the caustic).
  A first analysis reveals good agreement between UVES and synthetic spectra for the 
wings part of $\element{H}\alpha$ ($6558-6561.8$~\AA{} and $6563.8-6567.6$~\AA{}), whereas a clear discrepancy is 
observed in its core ($6561.8-6563.8$~\AA{}). We note that the chromosphere is not included in the PHOENIX 
calculations. Althought it should have a minor influence on the broadband limb darkenings, it will have an effect
on the core of lines like $\element{H} \alpha$. We therefore divided the analysis in two parts: the wings
and the core.
  In order to compare the observed equivalent widths of the $\element{H}\alpha$-line with the predictions from our
synthetic spectrum, we apply an overall scale factor to the equivalent width
($1.035$ for the wings part and $1.495$ for the core part of the line), so that the measurements derived from 
the fiducial model match the post-caustic observations. 

The observed and predicted temporal variations of the equivalent widths of
both the wings and the core of $\element{H}\alpha$  are plotted in the upper part of Fig.~\ref{fig:AetAcurve}. 
Note at the beginning a small short-term decrease in the model, when the leading limb hits
the caustic, followed by little variation while the source centre crosses, but then
 a marked change when the trailing limb exits the caustic.
   The spectral profile of the core is not well matched by the model spectrum as seen in Fig.~\ref{fig:ha};
both the width and the depth disagree. Furthermore, the differential variation
in flux between the centre and the limb is not well reproduced, even after the
scale factor has been applied. A better fit is obtained to the wings.
Both parts of the line show larger equivalent width variations than predicted
by the model.

\begin{figure}
\includegraphics[width=8.cm]{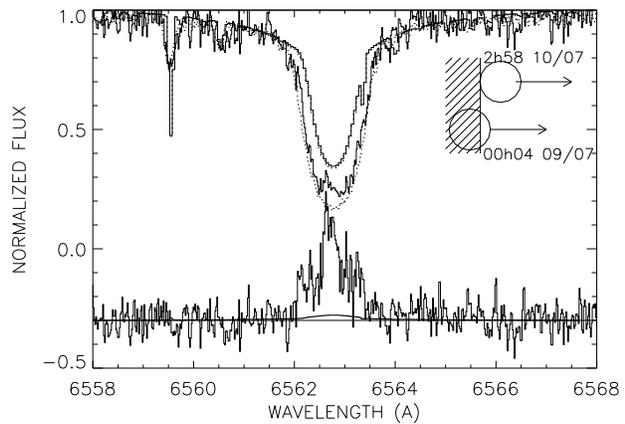}
\caption{ Same Legend as Fig.~\ref{fig:ha} but for the pair of spectra 9 July at UT 00h04.
and 10 July, UT 02h58. Notice the strong signature of the chromospheric $\element{H} \alpha$ emission.    }
\label{fig:ha2}
\end{figure}

  Last but not least, in the spectrum corresponding to 10 July, UT 02h58 shown
in Fig.~\ref{fig:ha2}, we clearly note a strong deviation in the 
core of the line but a much smaller one in the wing. For the purpose of the following 
discussion, we define the effective photospheric radius of the star as that at which the 
R band intensity has fallen to 5 \% of the central intensity in R. Using the fit 
done with limb darkening derived from PHOENIX model, we infer that the spectrum started
when the caustic was at a fractional radius $\rho=0.996$ and ended at $\rho=1.015$ while 
with a linear limb-darkening law, the spectrum started at a fractional radius $\rho=1.007$
and ended at $\rho=1.026$.

In both cases, it is clear that the spectrum was taken when the caustic was outside the photosphere in R.
There is a clear emission peak in the core of the $\element{H} \alpha$ absorption line, 
and this can only exist provided there is a temperature inversion in the atmosphere (i.e. a chromosphere). 
The doppler shift of the emission core is consistent with material moving outward through the 
giant source star's chromosphere ($\element{H} \alpha$ emission line) with a radial velocity 
of $\approx$ 7 $\rm{kms}^{-1}$, the signature being amplified at the very end of the caustic exit.

%
\begin{acknowledgements}
%
For this particular VLT follow up exercise, we are expressing our highest
gratitude to the ESO staff at Paranal and La Silla whose flexibility and efficiency
was absolutely vital to the success of the observations.
We are very grateful to OGLE-III for their Early Warning System (EWS), to the observatories that support our
science (European Southern Observatory, Canopus, CTIO, Perth, SAAO)
via the generous allocations of time that make this work possible. 
 We thank Andy Gould
and Dimitar Sasselov for fruitful discussions. The
operation of Canopus Observatory is in part supported by the financial
contribution from David Warren, and the Danish telescope at La Silla
is operated by IJAF financed by SNF.  JPB acknowledges
financial support via award of the ``Action Th\'{e}matique Innovante''
INSU/CNRS. MD acknowledges postdoctoral support on the PPARC rolling grant
PPA/G/O/2001/00475.
\end{acknowledgements}
%
%
\bibliographystyle{aa}
\bibliography{Ga073}
\end{document}